\def\section{\@startsection {section}{1}{\z@}{-26pt plus -1ex minus
    -.2ex}{13pt plus .2ex}{\large \bf}}
\def\subsection{\@startsection{subsection}{2}{\z@}{-13pt plus -1ex minus
   -.2ex}{13pt plus .2ex}{\large \rm}}
\def\subsubsection{\@startsection{subsubsection}{3}{\z@}{-13pt plus
  -1ex minus -.2ex}{-9pt plus .2ex}{\large \em}}
\def\paragraph{\@startsection
     {paragraph}{4}{\z@}{3.25ex plus 1ex minus .2ex}{-1em}{\large\bf}}
\def\subparagraph{\@startsection
     {subparagraph}{4}{\parindent}{3.25ex plus 1ex minus
     .2ex}{-1em}{\large\bf}}

\newcommand{\heading}[1]{\vspace*{15mm}
{\Large\begin{center} {\bf{#1}} \end{center}}}
\renewcommand{\author}[3]{\vspace{5mm}
\begin{center}
{\normalsize \rm #1}\\    
{\normalsize \it #2}\\    
{\normalsize \it #3}\\    
\vspace{0.65cm}\framebox[3.8truecm]{\rule[-1.9cm]{0.cm}{3.8truecm}}
\vspace{0.35cm}\end{center} }

\def\lsim{~\rlap{$<$}{\lower 1.0ex\hbox{$\sim$}}}
\def\gsim{~\rlap{$>$}{\lower 1.0ex\hbox{$\sim$}}}

\newcommand{\mn}{{\em Mon. Not. R. astr. Soc}}
\newcommand{\apj}{{\em Astrophys. J.}}

\documentstyle[11pt]{article}
\centerline{Talk presented at the 9th IAP Meeting: ``Cosmic Velocity Fields'',
Paris, July '93}

\def\gtorder{\mathrel{\raise.3ex\hbox{$>$}\mkern-14mu
             \lower0.6ex\hbox{$\sim$}}}
\def\ltorder{\mathrel{\raise.3ex\hbox{$<$}\mkern-14mu
             \lower0.6ex\hbox{$\sim$}}}

\def \lsim {\ltorder}
\def \gsim {\gtorder}
\def \hide#1{}

\begin{document}
\heading{TESTING MODELS FOR STRUCTURE FORMATION}

\author{N. KAISER}
       {CIAR Cosmology Programme}
       {CITA, University of Toronto, Toronto M5S 1A7}

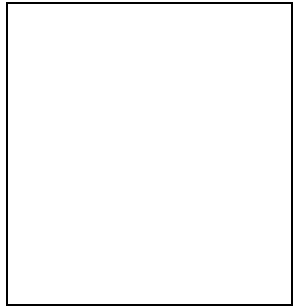
\begin{abstract}{\baselineskip 0.4cm
I review a number of tests of theories for structure formation.
Large-scale flows and IRAS galaxies indicate a high density parameter
$\Omega \simeq 1$, in accord with inflationary predictions,
but it is not clear
how this meshes with the uniformly low values obtained from
virial analysis on scales $\sim$1Mpc.  Gravitational distortion
of faint galaxies behind clusters allows one to construct maps
of the mass surface density, and this should shed
some light on the large vs small-scale $\Omega$ discrepancy.
Power spectrum analysis reveals too red a spectrum (compared to
standard CDM) on scales $\lambda \sim 10-100$ $h^{-1}$Mpc, but the gaussian
fluctuation hypothesis appears to be in good shape.
These results suggest that the
problem for CDM lies not in the very early universe ---
the inflationary predictions of $\Omega = 1$ and
gaussianity both seem to be OK; furthermore, the COBE result severely restricts
modifications such as tilting the primordial spectrum --- but in the assumed
matter content. The power spectrum problem can be solved
by invoking a cocktail of mixed dark matter. However, if gravitational
lensing fails to reveal extended dark mass around clusters then we
may be forced to explore more radical possibilities for the
dark matter.
} \end{abstract}

\section{Introduction}
Current theories for structure formation, as exemplified by the
standard cold dark matter model,
are based on two quite distinct pieces of physics:
First there is the inflationary phase in the very early universe
when $H$ is on the order of the Planck or GUT mass, and when it is assumed
i) that the universe is dominated by a nearly spatially constant
scalar field $\phi_0$ which is rolling slowly down the potential $V(\phi)$
and which drives inflation, thus preparing the universe with $\Omega$ very
close
to unity and
ii) that there are zero-point fluctuations of the field $\delta \phi$
which, at horizon crossing, leave their imprint as a very nearly
scale-invariant
spectrum of gaussian density fluctuations $\delta \rho/\rho$.
The second piece of physics describes the modification of the
initially featureless spectrum as the universe passes from radiation
to matter domination: $P(k) \rightarrow T^2(k) P(k)$, and where the
transfer function $T(k)$ depends critically on the dark matter content.

I will describe here some tests for the various predictions listed
above.  These tests fail miserably to disprove the predictions of
the inflationary scenario --- dynamics of large scale structure suggests
$\Omega \simeq 1$ and the fluctuations appear consistent with the
gaussian hypothesis --- but the shape of the spectrum on scales
10-100 Mpc suggests we may need to modify our assumptions
about the dark matter.
\vfill\eject

\section{Tests for $\Omega$}

\subsection{$\Omega$ from large-scale flows}
There is a growing consensus that i) there is a positive correlation between
large-scale peculiar motions and the gravity inferred from the
galaxy distribution (as would be expected under gravitational instability) and
ii) that the relative amplitude of the peculiar motions and the (mainly IRAS)
galaxy density contrast implies a high density parameter $\Omega \simeq 1$ if
these galaxies fairly trace the mass.

The POTENT comparison with IRAS galaxies has been discussed elsewhere
in these proceedings.  These results are in broad agreement with the
$\Omega$ inferred from the QDOT redshift survey \cite{QDOT},
which in turn reinforced the high $\Omega$ inferred from the angular dipole
moments (\cite{DIPOLE} and references therein).
While there are relative advantages and disadvantages of all these
approaches, there is generally good agreement (though similar efforts
using optical galaxies tend to give lower $\Omega$, perhaps implying that
the optical galaxies are somewhat positively biased).

The QDOT approach was to estimate the smoothed gravity from the 1-in-6 IRAS
0.6Jy redshift survey, correct this for redshift space distortion effect,
and then compare this with the peculiar velocity smoothed on the same
scale.  This avoids the inhomogeneous Malmquist bias effect
which enhances the density contrast in the POTENT maps,
but at the price of sacrificing the locality of the POTENT
velocity-divergence comparison.  We tried to minimise the effect of structures
beyond the survey volume by
using predicted velocities relative to the local group.
A comparison with the motions of a set of clusters with fairly precise
distant estimates \cite{FRENK} is shown in figure 1.

\begin{figure}[h]
\caption{Comparison of the smoothed velocity predicted from the
IRAS QDOT survey (assuming $b = \Omega = 1$) with the observed motions of a set
of
clusters from Frenk {\it et al.\/}, 1993.}
\end{figure}

This shows (somewhat more convincingly than in our previously
published scatter plot) that if IRAS
predicts that a cluster should be moving towards or away from us then it will
with great probability be found to be moving in the expected sense.  For
$\Omega = 1$ the points should scatter about a line of slope 45$^\circ$, which
looks about right. The formal error on $\Omega$ in this and other studies is
typically 30\%.  This is of course questionable, but it is clear at least
that a density parameter equal to the baryonic density predicted
by big bang nucleosynthesis $\Omega_{\rm BBN} \simeq 0.015 h^{-2}$
\cite{BBN} would
be very hard to reconcile with this result --- we would need
a strong antibias --- so we have evidence
here that the dark matter is non-baryonic.

\subsection{$\Omega$ from lensing}
The large $\Omega$ implied from large-scale ($\gtorder 30 h^{-1}{\rm Mpc}$)
studies stands in contrast to the low mass-to-light ratios $M/L \sim
100-300h$ (implying $\Omega \ltorder 0.2$) on scales
$\sim 1 {\rm Mpc}$ inferred from virial analysis
of clusters and from the cosmic virial theorem (Davis and Peebles, \cite{DP}).
This may be partly reflecting a relative bias of IRAS and optical galaxies,
but this most probably also reflects a real dependence on scale.
An attractive possibility is that the small scale estimates are biased
downwards because the galaxies are more concentrated than the dark
matter in clusters: i.e.\ clusters, like galaxies, have dark halos.

We can test this hypothesis with gravitational  lensing: Dark mass in the
outskirts of clusters will produce small, but coherent, distortions of the
background galaxies.  The effect is weak at large radii (the distortion
falls roughly like the projected surface mass density), and this is where we
are most interested in probing the mass.  However, the potential sensitivity
improves at large angle due to the increase in the number of background
galaxies.  For a space density profile $\rho \propto 1/r^2$, the signal
to noise should be independent of radius.  The effect is clearly seen in
the inner parts of clusters so a $\rho \propto 1/r^2$ continuation to
large scales should also be clearly seen (and steeper profiles could be
measured
statistically by stacking the results from a number a clusters).

Here's how it works.  One measures quadrupole moments of the faint
background galaxies and constructs ellipticity or polarization parameters
$e_1,e_2$ which describe stretching of the galaxy images along the axes
and diagonals respectively.  The galaxy ellipticities provide
noisy estimates of a gravitationally induced ellipticity pattern which
is just the convolution of the projected mass density field with
the pattern from a point mass: $e_i \propto \chi_i(\vec \theta) = \{\cos 2\phi,
\sin
2\phi\}/ |\theta|^2$:
\begin{equation}
\hide{$$}
e_i = \chi_i \otimes \Sigma
\label{eq:E1}
\hide{$$}
\end{equation}
from which we can reconstruct the projected surface density $\Sigma(\vec
\theta)$ (measured in units of the critical density).  In fact
\cite{KS} the optimal inversion is
just the convolution and contraction of the observed ellipticity with the same
kernel $\chi_i$:
\begin{equation}
\hide{$$}
\Sigma = -\chi_i \otimes e_i
\label{eq:E2}
\hide{$$}
\end{equation}
The, assumed random, intrinsic ellipticities of the background galaxies
provides a source of noise, and this means that one must smooth the
dark matter map in some way.
A further complication is that the method cannot determine the average
surface density so in practice this means that about half of the
data are used to set the baseline surface density.

This probe of DM was pioneered by Tyson, Valdes and Wenk \cite{TVW} with
their study of A1689, though using a different convolving
kernel.  More recently Ellis
and Smail \cite{SMAIL} have obtained interesting results for Cl1455-22
using (\ref{eq:E2}); their dark matter map shows, in addition to the
main cluster, a second blob, which is also seen as an enhancement in
the ROSAT X-ray image.
Fahlman, Squires, Woods and myself \cite{FAHLMAN} have applied this technique
to
CFH images of A2218 and MS1224.

\begin{figure}
\caption{Projected mass map for MS1224. The total field here is
$\sim 12'$ on a side and is composed of four $7' \times 7'$ CCD images.
Only contours above 1-sigma of estimated noise level are shown.
At the cluster redshift of 0.33 the box side corresponds to a comoving
scale of 3$h^{-1}$Mpc.}
\end{figure}

No prominent giant arcs are seen in this cluster, yet the
mass reconstruction clearly shows the dark mass in the centre of this
not exceptionally massive cluster (the central velocity dispersion
is 800 km/s).
This is is the largest field studied so far, with about 1000 background
galaxies measurable, and is
the first to allow any view of the dark mass beyond about 1/2 Mpc.
Unfortunately (for inflation), our preliminary analysis indicates
than the DM is, if anything, more concentrated than the galaxies in this
cluster.

This is a very promising probe of the DM which will surely take off
with the advent of large format CCD mosaics.  A failure to detect an
increase of M/L with radius would be very interesting, and would
point to one of two radical conclusions:

\noindent
1) $\Omega$ is really low and nearly all of the large-scale structure
studies reported at this meeting are incorrect --- heaven forbid!

\noindent
2) $\Omega$ is really unity but for some reason the dark mass
cannot cluster on scales less than tens of Mpc.

\section{Power Spectrum and Gaussianity}
At a qualitative level it is a pleasant prediction of models like CDM
that the primordial spectrum should break from the primordial $n = 1$
spectrum at the scale of the horizon at $z_{\rm eq}$ to something flatter.
Several studies though have shown that at a more quantitative statistical
level the break occurs at too large a scale; to make standard CDM match
what is seen would require $\Omega h \simeq 0.2-0.25$ which, with $\Omega = 1$,
is an unacceptable Hubble parameter.
The first really strong evidence for extra large-scale power came from
the APM survey\cite{APM}.  Several groups have now performed
power-spectrum analysis on redshift surveys.  The spectrum extracted from the
QDOT redshift survey by Feldman {\it et al.\/} \cite{FELDMAN} is shown in
figure
3.

\begin{figure}[h]
\caption{Power spectrum of QDOT redshift survey.}
\end{figure}

This survey, being deep and of $\sim 3\pi$ steradians is particularly well
suited
to studying $P(k)$ at large scales.  A new feature of this analysis is that
the statistical uncertainty is calculated analytically under the assumption
that
the large-scale fluctuations are gaussian, and this allowed us to optimise the
weighting scheme.
The survey provides a fairly accurate $P(k)$ at wavelengths $\lambda
\sim 30-150 h^{-1}{\rm Mpc}$ where linear theory should be accurate,
and yet the spectrum clearly has the wrong shape for standard CDM.
Figure 3 also shows that as a phenomenological fix, the 70/30 cold/hot mixed
dark matter model \cite{KLYPIN} would fit the data quite well.

The strength of large-scale power is $P \simeq 10^4 (h^{-1} {\rm Mpc})^3$
at $k \simeq 0.05 h / {\rm Mpc}$ ($\lambda \simeq 120  h^{-1} {\rm Mpc}$).
It it perhaps more meaningful to express this as a contribution
to the variance:
\begin{equation}
\hide{$$}
\langle \sigma^2 \rangle = \int {d^3 k\over (2\pi)^3} P(k)
= (0.25)^2 \int d\ln k\; (k/0.05)^3 (P(k) / 10^4)
\label{eq:P1}
\hide{$$}
\end{equation}
So this level of power corresponds to a rms density fluctuation of about 25\%.

It is also interesting to relate this power to other probes of large-scale
structure: peculiar velocity; microwave anisotropy and
coherent gravitational distortions.
A quantitative prediction for the rms signal in a given experiment
requires one to calculate the `window function' which determines the
sensitivity as a function of spatial wavelength.  This is usually
straightforward but is experiment dependent.  One can get a good idea
of how much signal might be detected by estimating the contribution
per log interval of wave number as in (\ref{eq:P1}).
Assuming $b_{\rm IRAS}
\simeq 1$ and $\Omega = 1$ the observed power would drive
streaming motions of rms amplitude $\simeq$ 500 km/s:
\begin{equation}
\hide{$$}
\langle v^2 \rangle = \int {d^3 k\over (2\pi)^3}
{H^2 \over k^2} P(k)
= (500 {\rm km\;s}^{-1})^2 \int d\ln k\; (k/0.05) (P(k) / 10^4)
\label{eq:P2}
\hide{$$}
\end{equation}
so there should be substantial motions even on these rather
large scales.

Relating $P(k)$ to the microwave anisotropy
is more model dependent. Naively applying the Sachs-Wolfe result
gives an rms anisotropy of $\delta T/T \simeq 5 \times 10^{-6}$:
\begin{equation}
\hide{$$}
\langle (\Delta T/T)^2 \rangle = {1\over 4} \int {d^3 k\over (2\pi)^3}
{H^4 \over k^4} P(k)
= (5 \times 10^{-6})^2 \int d\ln k\; (k/0.05)^{-1}
 (P(k) / 10^4)
\label{eq:P3}
\hide{$$}
\end{equation}
The Sachs Wolfe result is not, however, directly applicable on these
scales, but what this
means is that extrapolating to the $\sim 1000 h^{-1} {\rm Mpc}$
scales probed by COBE with a scale-invariant $P(k) \propto k$ would give
roughly
what is observed.

Finally, these density fluctuations would cause distortions
of distant galaxies coherent over degree scales at about the
1\% level:
\begin{equation}
\hide{$$}
\langle e^2 \rangle = {12 \pi w_s^3\over 5}
\int {d^3 k\over (2\pi)^3}
{H \over k} P(k)
= (9 \times 10^{-3})^2 (w_s / 0.3)^3 \int d\ln k\; (k/0.05)^{2}
 (P(k) / 10^4)
\label{eq:P4}
\hide{$$}
\end{equation}
with $w_s \equiv 1 - (1 + z_s)^{-1/2}$. Hudon {\it et al.\/} \cite{HUDON}
have obtained limits at about this level, and the prospects for improving
sensitivity here are good.

It is apparent from the foregoing equations that the power spectrum
provides a very convenient vehicle for comparison with other
probes of large-scale structure.  The comparison can be made precise simply
by inserting the appropriate window function (or rather it's fourier
transform squared) in the above equations.  A further important advantage
of power spectrum over correlation function is that the former effectively
diagonalises the error matrix and this greatly facilitates the
calculations of likelihood ratios for competing theories, which is after all
surely the ultimate goal of these studies.

We have also used the power spectrum to test the gaussian hypothesis.
While the mean value of the power is void of information regarding
gaussianity, the fluctuations about the mean can reveal some
traces of primordial non-gaussianity.  We have estimated the 1-point
distribution of the power.  According to the gaussian hypothesis this
should be exponential, and the observed distribution is in very good
agreement with this.

\begin{figure}
\caption{1-point distribution of the power from the
IRAS-QDOT redshift survey}
\end{figure}

We have also looked at the two point function of the power spectrum:
$\chi(\delta k; k) = \langle P(\vec k) P(\vec k + \vec {\delta k})
\rangle$.  Under the gaussian hypothesis this is a bell-shaped function
localised
around $\delta k = 0$ with width $\sim 1/D$, the inverse effective depth of the
survey. An excess width to the two-point function can arise in models where
one has one gaussian random field with strong low-spatial frequency
power amplitude modulating a second independent gaussian field
with strong high frequency fluctuations; crudely speaking $\chi(\delta k; k)$
measures the extent to which modes of wavenumber $k$ are being
modulated by modes of wavenumber $\delta k$. Such a model has been
suggested by Peebles \cite{PEEBLES} as phenomenologically attractive, and this
behaviour can arise in certain inflationary models (Bardeen,
personal communication).
Again, in the data, we see no evidence for non-gaussian behaviour.

\section{Alternatives to Collisionless Cold Dark Matter?}
One cannot help but be impressed by how close the standard CDM model
comes to explaining the observed structure; the halos produced
can arguably account nicely for galaxies and clusters, the
texture of large scale structure with its sheets and
filamentary appearance is a very good match at the visual level
to observed galaxy clustering. Moreover,
the detection of MBR anisotropy by COBE at about the predicted
level provides very strong
support to the general idea that present day structure can indeed be
traced back to the very early universe rather than being the outcome
of astrophysical processes.  The problems with the model only appear
at a rather detailed statistical level of comparison.  One problem
stands out very clearly; the shape of the power spectrum at
large scales.  To this we should add the Hubble constant problem
and --- though these seem less convincing as yet --- the baryon
content of clusters and the large-scale vs small-scale $\Omega$
discrepancy.

The qualitative success of the CDM model suggest that what is
needed is tinkering rather than a complete rethink.
By this I mean that one should consider modifications to either
the very early universe aspects or the medieval universe, but probably
not both.  As the data seem quite compatible with the inflationary
expectations as regards $\Omega$ and gaussianity,
and since the possibility to modify the model by e.g.\ tilting
the initial spectrum is now highly constrained by COBE,
it would seem more promising to explore alteratives
to the simple, if aesthetically pleasing, assumption that the dark matter is
cold and collisionless.

One simple fix for the $P(k)$ problem is to invoke a cocktail of
hot and cold dark matter, and
a 70/30 cold/hot mix \cite{KLYPIN} gives an acceptible power spectrum.
However, having to invoke two different types of dark matter
is clearly a demerit point for this theory as compared to
standard CDM or HDM.  A significant development in this regard
is the proposal of Madsen \cite{MADSEN}, who argued that
(non-radiative)
decays of a massive neutrino while still relativistic could
naturally produce bosonic decay products with a substantial
fraction of the particles in a bose condensate and that this
could therefore create mixed dark matter with the cold and
hot particles being one and the same species.  Stimulated by this,
but not completely convinced by Madsen's arguments which involved
some questionable assumptions, Malaney, Starkman and myself
\cite{KMS} have formulated and solved the Boltzmann equations
for such a decay process.  We have found that a bimodal momentum
distribution for the bosonic decay products can indeed result,
though by a rather different process than the thermalisation
envisaged by Madsen.  What we found was that under certain conditions
there is a runaway process of decays to very low momentum
bosons or `neutrino lasing'.  This is quite different from the
more often considered case of non-relativistic decays where the
decay products are extremely hot.  For relativistic decays,
there is the possibility to produce rather cold products
if these are emitted in the opposite direction to the motion
of the decaying particle.  As the phase space at low momenta
is small, the occupation number rapidly becomes
large.  This stimulates more low momentum decays
and very soon essentially all of the decays are to very
cold bosons.

The lasing happens when the decaying particle is relativistic.
This is followed by a second phase of decays when the massive neutrino
goes non-relativistic, and these result in hot bosons.
The result, if the boson has a mass of a few tens of eV, is
mixed dark matter.  The details of the model predictions
depend on such features as whether the boson is its own antiparticle.
The specific model considered in \cite{KMS}, for instance, produced a
cold-fraction of 37-50\%, rather less than the preferred 70\%
fraction.  However, the hot particles in this model came out
to be somewhat cooler than in the standard {\it ad hoc\/.} model \cite{KLYPIN},
so this may well still be viable. If not, then there are other
more or less
plausible ways to increase the cold fraction within this scheme.

Mixing collisionless dark matter is only one possibility.
Another interesting avenue is to drop the assumption that the
dark matter is collisionless. After all, we only know that the
dark matter does not interact very much with photons and baryons;
whether the dark matter is self-interacting is unknown.
Carlson, Machacek and Hall \cite{CH} have explored the possibility that
the DM self-interacts and that number changing interactions
are effective.  This yeilds the interesting result that after
becoming non-relativistic  the
dark matter cools more slowly than would be the case for
collisionless matter.  This might seem an attractive way
to solve the ``$\Omega$-discrepancy'' (and the closely related problem of
the baryonic fraction in clusters) as keeping the dark matter
warm would effectively stop it clustering on small scales.
However, it seems to be rather difficult in this scheme to arrange
matters so that the matter can cluster on tens of Mpc scales.
A second problem with this scenario is that there is strong
suppression of the growth factor on small scales, so it may be
difficult to make galaxies.  Another possibility is to assume
that the dark matter is extremely light (e.g.\ \cite{HR}) with
a mass on the order of $10^{-25}$eV.  Such fields have been widely
considered in the context of late-time phase transitions, and
have the very interesting property that the particles are
forbidden from clustering on small scales because of the
wave-mechanical uncertainty principle --- the particles
cannot be bound within a structure with velocity
dispersion $\sigma$ if the de Broglie wavelength $\sim h / m \sigma$
is smaller than the size of the system.  There is also the
possibility that the dark matter is collisional, but that
number changing reactions are ineffective; if so
the difference in clustering properties from the collisionless case
would be more subtle, but potentially interesting nonetheless.

\vfill
\end{document}